# Solving and refining novel thin film phases using Cu X-ray radiation: the epitaxy-induced CuMnAs tetragonal phase

P. Wadley[1,6], A. Crespi[2], J. Gazquez[2], M.A. Roldan[3,4], P. Garcia[5], V. Novak[1], R. Campion[6], T. Jungwirth[1,6], C. Rinaldi[8], X. Marti[1,7,9,*], V. Holy[7], C. Frontera[2], J. Rius[2]

[1] Institute of Physics ASCR, v.v.i. Cukrovarnická, Praha, Czech Republic
[2] Institut de Ciència de Materials de Barcelona, ICMAB-CSIC, Bellaterra, Spain
[3] Departamento de Física Aplicada III, Universidad Compluense de Madrid, Madrid, Spain
[4] Materials Science & Technology Division, Oak Ridge National Laboratory, USA
[5] Centre Investigació Nanociència Nanotecnologia, CIN2, Bellaterra, Spain
[6] School of Physics and Astronomy, University of Nottingham, United Kingdom
[7] Faculty of Mathematics and Physics, Charles University in Prague, Czech Republic
[8] LNESS – Dipartimento di Fisica Politecnico di Milano, Como (Italy)
[9] Department of Physics, University of California, Berkeley, USA

We present a combined experimental and computational method which enables the precise determination of the atomic positions in a thin film using CuK$\alpha$ radiation, only. The capabilities of this technique surpass simple structure refinement and allow solving unknown phases stabilized by substrate-induced stress. We derive the appropriate corrections to transform the measured integrated intensities into structure factors. Data collection was performed entirely on routinely available laboratory diffractometers (CuK$\alpha$ radiation); the subsequent analysis was carried out by single-crystal direct methods ($\delta$ recycling procedure) followed by the least-squares refinement of the structural parameters of the unit cell content. We selected an epitaxial thin film of CuMnAs grown on top of a GaAs substrate, which formed a crystal structure with tetragonal symmetry, differing from the bulk material which is orthorhombic. Here we demonstrate the new tetragonal form of epitaxial CuMnAs grown on GaAs substrate and present consistent high-resolution scanning transmission electron microscopy and stoichiometry analyses.

* xaviermarti@berkeley.edu

## I. Introduction

Subtle modifications of the unit cell topology can lead to dramatic changes in the magnetic, dielectric, optic, chemical, etc., properties of materials. From this perspective, one of the major advantages of thin film growth is the ability to fine tune lattice parameters, angles and distances in the unit cell to obtain enhanced or completely new functionalities. For bulk single crystals and powder samples well established methods of structural refinement exist, which are routinely performed with unrivalled accuracy. However, in the case of epitaxial thin films, the application of such a complete set of tools to data acquired using conventional X-ray equipments, is still a growing field [1], despite a substantial progress in X-ray large facilities (see Ref. [2] and references therein). Yet the majority laboratory analyses is restricted to the accurate determination of the lattice parameters, and does not provide critical information such as bonding angles (important in magnetic interactions) or bond lengths (for instance, connected to electrical transport anisotropies).

To obtain a complete description of the unit cell of a highly textured material, two fundamental pieces of information are required: firstly, the lattice parameters and, secondly, the relative intensities of as many diffraction peaks as possible – including systematic extinctions. While a conventional high-resolution diffraction experiment can accurately determine the lattice parameters from a reduced set of reflections, but it is unable to provide an exhaustive list of relative peak intensities in a reasonable time frame, owing to the small diffracting volume. Moreover, the high-resolution (HR) coplanar set-up for thin epilayers obscures several reflections due to the shadowing effect of the sample holder. To overcome these issues, two-dimensional plate detectors placed at short distances from the thin film sample are a time-efficient way to collect the integrated intensities but at the cost of lowering the resolution. By combining the data from both experimental set-ups it is possible to perform the complete study of the thin layer in a precise and time-efficient manner.

Here we report on the unit cell solution and refinement of a thin layer material with no a priori knowledge of the crystal structure. We present explicitly the formulae to compensate the measured intensities when using a two dimensional plate-detector in the conventional z-axis geometry. By considering their angular coordinates, the corresponding intensities are integrated and subsequently processed by single-crystal direct methods ($\delta$ recycling) to propose a structure model which is later optimized. The experimental part only requires CuK$\alpha$ -radiation sources and two different detectors to collect the integrated intensities and to determine the lattice parameters separately. To demonstrate our methodology, we have selected the situation in which a MBE grown thin film crystallizes in a phase which is different from the phase of the bulk compound, despite having the same stoichiometry.

The paper is organized as follows: In Section II we describe the sample preparation, the experimental set-up and the measurement strategy. The corrections for integrating the peak intensities according to their angular location are derived in Section III. The application of $\delta$ recycling and the refinement of the unit cell contents are discussed in Section IV. Section V summarizes the structural analyses results which are compared to those obtained from high resolution scanning transmission electron microscopy (HR-STEM) and stoichiometry analyses.

## II. Sample preparation and experimental set-up

The CuMnAs thin layer, which we used to validate this study, was grown by molecular beam epitaxy on GaAs(001) substrates. Details of the sample preparation can be found elsewhere [3]. A 5 x 2 mm sample was cut from the original wafer for use in the present study. The thickness of the CuMnAs layer is 130(3) nm according to X-ray reflectivity (not shown). The relative intensities have been collected on a Bruker D8 diffractometer equipped with an X-ray mirror, a double pinhole on the incident beam side and a GADDS two dimensional detector on the diffracted beam side, located 14 cm from the sample. Details and illustration of the set-up are exhaustively covered in Ref. [4]. In this set-up the detector can rotate only around the scattering angle $2\theta$, covering on the detector surface $\Delta 2\theta \sim 30$ degrees within the scattering plane and $\Delta\chi \sim 10$ degrees away from the scattering plane. Azimuthal rotations and rocking of the sample enable collection of the reciprocal space peaks with the detector. An exhaustive description of the employed angles can be found in Ref. [4]. In this paper we follow the same angle and sign conventions contained therein. By means of a Cu mask we restricted the active area of the detector to 3 cm in diameter to avoid the simultaneous counting of two or more very intense substrate peaks. High resolution X-ray diffraction experiments have been performed using a Panalytical X'Pert Material Research Diffractometer, equipped with an X-ray mirror and a Bartels monochromator in the incident beam side, and a PixCel linear detector on the diffracted beam side. Both set-ups used Cu-anode tubes.

From the Reflection High Energy Electron Diffraction (RHEED) patterns, collected during the growth, the sample is expected to be in-plane textured. Accordingly, we explored the vicinity of the substrate peaks expecting to detect additional reflections. We collected several pole figures around the arbitrarily chosen Ga(202) reflections and within the $\Delta 2\theta \sim 30$ degrees and $\Delta\chi \sim 10$ degrees. We observed several diffraction peaks that could not be ascribed to GaAs (Fig. 1a). The arrangement of the peaks suggested a tetragonal symmetry with only very weak traces of potentially orthorhombic ordered material. This is at odds with the orthorhombic crystal structure of bulk CuMnAs [5, 6]. We explored in detail one of the reflections using the HR set-up to accurately determine the lattice parameters of the tentative tetragonal structure. As shown in Fig. 1b, we obtained lattice parameters a = 3.820(5) Å and c = 6.318(5) Å. These lattice parameters were used to determine the angular coordinates in the z-axis geometry of all the (*hkl*) peaks, assuming that the lattice angles are equal to 90 degrees. A batch data collection program was used for the subsequent intensities measurement. The observed intensities are listed in Table I. Systematic absences correspond to to $l=0$ [(*hk*0) with $h+k$ odd and (0*k*0) with *k* odd]. This set of extinctions (see for instance Fig. 1c the absence of x-ray diffraction of (010) which in contrast has been observed in neutron diffraction [3]) suggests the space group P4/nmm (no. 129) which is compatible with all observations. Consistently, our electrical measurements, X-ray absorption spectroscopies and RHEED patterns also presented a 4-fold symmetry [3]. Our conclusion is that the CuMnAs epilayer is adopting the tetragonal structure of bulk $Mn_2As$ with the same space group and very similar lattice parameters [7,8]. Recent crystal growth experiments have demonstrated that bulk CuMnAs can also be stabilized in the tetragonal phase with space group and lattice parameters very similar to our findings in analogous thin epilayers [9].

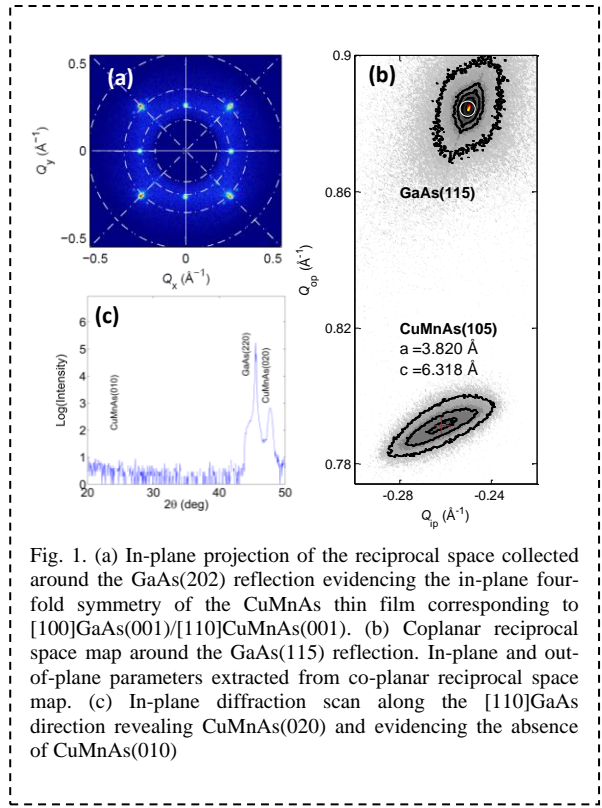

Fig. 1. (a) In-plane projection of the reciprocal space collected around the GaAs(202) reflection evidencing the in-plane four-fold symmetry of the CuMnAs thin film corresponding to [100]GaAs(001)/[110]CuMnAs(001). (b) Coplanar reciprocal space map around the GaAs(115) reflection. In-plane and out-of-plane parameters extracted from co-planar reciprocal space map. (c) In-plane diffraction scan along the [110]GaAs direction revealing CuMnAs(020) and evidencing the absence of CuMnAs(010)

Errors caused by sample misalignment have been quantified by comparing the integrated intensities of four symmetry equivalent reflections of GaAs(202). We have found 11% intensity spread and this value has been taken into considered as uncertainty for the computations.

### III. Corrections of peak intensities

To collect the intensity for a given reflection we oriented the sample and the center of the detector according to the (*hkl*) reciprocal space positions using the conventions in Ref. [4]. In the following we will distinguish two types of scans: φ and ω-scans. In the φ scans, we rotated the sample on the sample holder azimuthally while we continuously append the instantaneous intensity incident on the detector into a single two-dimensional image. The process is equivalent to the diffraction peak travelling across a plane in the reciprocal space containing the vectors of the sample texture and one in-plane vector depending on the actual azimuth. In the ω-scans we rock the sample while keeping the plate detector fixed. In this case, the excursion of the diffraction peak across the detector has no simple analogue in reciprocal space [4]. In all of the collected images we were able to mask completely, or nearly completely, the nearby substrate peaks by using the aforementioned mask with 3 cm diameter aperture. The background contribution has been calculated by fitting a surface with polynomials of first order to the surrounding region of the film. The integrated intensities have been obtained by integrating the two-dimensional Gaussian fit of the data. Data from the most intense collected reflections are summarized in Table I.

Table I. Relative integrated intensities for each (*hkl*) reflection located at angular positions $2\theta$, $\psi$, and $\omega$. Angles are shown in degrees.

| *h* | *k* | *l* | $2\theta$ | $\psi$ | $\omega$ | $I^M_{hkl}$ |
|---|---|---|---|---|---|---|
| 0 | 0 | 1 | 16.00 | 0.00 | 8.00 | 43.40 |
| 0 | 0 | 2 | 28.00 | 0.00 | 14.00 | 14.05 |
| 0 | 0 | 3 | 43.00 | 0.00 | 21.50 | 44.78 |
| 0 | 0 | 4 | 58.50 | 0.00 | 29.25 | 10.40 |
| 0 | 1 | 6 | 99.12 | 15.45 | 47.19 | 0.35 |
| 2 | 2 | 4 | 97.36 | 49.54 | 29.27 | 2.52 |
| 2 | 3 | 1 | 94.95 | 80.51 | 7.09 | 2.05 |
| 1 | 3 | 3 | 94.68 | 60.23 | 21.51 | 0.60 |
| 0 | 3 | 3 | 90.49 | 58.90 | 21.52 | 2.19 |
| 1 | 3 | 2 | 86.11 | 70.50 | 13.24 | 16.61 |
| 2 | 2 | 3 | 85.35 | 57.38 | 21.52 | 5.82 |
| 1 | 1 | 5 | 84.80 | 25.13 | 37.68 | 3.90 |
| 1 | 2 | 4 | 83.36 | 42.83 | 29.28 | 0.09 |
| 0 | 3 | 2 | 81.90 | 68.10 | 14.15 | 0.24 |
| 1 | 3 | 1 | 80.94 | 79.21 | 7.02 | 9.82 |
| 0 | 1 | 5 | 80.18 | 18.35 | 37.68 | 8.71 |
| 0 | 2 | 4 | 78.87 | 39.66 | 29.27 | 4.86 |
| 2 | 2 | 2 | 76.68 | 67.90 | 13.56 | 1.17 |
| 0 | 3 | 1 | 76.67 | 80.64 | 5.79 | 3.85 |
| 2 | 2 | 1 | 71.34 | 77.97 | 7.02 | 0.74 |
| 1 | 2 | 3 | 71.05 | 51.02 | 21.51 | 4.35 |
| 1 | 1 | 4 | 68.93 | 30.38 | 29.28 | 0.66 |
| 0 | 2 | 3 | 66.28 | 48.00 | 21.46 | 13.88 |
| 0 | 1 | 4 | 63.92 | 22.52 | 29.27 | 0.50 |
| 1 | 2 | 2 | 61.69 | 61.66 | 14.15 | 1.01 |
| 0 | 2 | 2 | 56.51 | 60.00 | 13.69 | 2.68 |
| 0 | 0 | 0 | 55.99 | 78.00 | 5.60 | 7.54 |
| 1 | 1 | 3 | 55.35 | 38.01 | 21.51 | 0.47 |
| 0 | 2 | 1 | 50.10 | 73.22 | 7.02 | 3.47 |
| 0 | 1 | 3 | 49.55 | 31.00 | 20.99 | 18.85 |
| 1 | 1 | 2 | 44.11 | 52.00 | 13.41 | 124.70 |
| 0 | 1 | 2 | 37.03 | 39.66 | 14.15 | 7.06 |
| 1 | 1 | 1 | 36.14 | 67.00 | 6.99 | 102.31 |
| 0 | 1 | 1 | 27.38 | 66.00 | 5.52 | 27.56 |

The integrated intensities, $I^M_{hkl}$, measured as explained previously, are not directly the square of the structure factor amplitudes $F_{hkl}^2$ for thin films. In order to obtain these quantities, which are needed for the structure solution and refinement, substantial corrections need to be taken into account. We applied Lorentz (L), polarization (P), irradiated volume (V), and absorption (A) corrections to the integrated intensities as follows:

$$F_{hkl}^2 = \frac{I_{hkl}^M L}{PVA} \quad \text{where:}$$

$$L = \begin{cases} \Delta\omega \dfrac{2\pi}{\lambda} \sin 2\theta & \text{(for an } \omega \text{ scan)} \\ \Delta\varphi \dfrac{2\pi}{\lambda} \sin 2\theta \sin \Psi & \text{(for a } \varphi \text{ scan)} \end{cases}$$

$$P = \frac{1 + \cos^2 2\theta}{2}$$

$$V = S \cdot t \frac{1}{\sin \gamma}$$

$$A = \frac{1}{\mu t \left(\dfrac{1}{\sin \gamma} + \dfrac{1}{\sin(2\theta - \gamma)}\right)} \left\{ 1 - \exp\left[-\mu t \left(\dfrac{1}{\sin \gamma} + \dfrac{1}{\sin(2\theta - \gamma)}\right)\right]\right\}$$

Δω and Δφ are the amount of corresponding angle rotated along the scan, λ is the wavelength, t is the film thickness, S the section of the incident beam, γ the incident angle, and μ the absorption coefficient. The derivation of these expressions is detailed below.

**III.1 Lorentz-equivalent correction**

A diffraction peak is spread around its theoretical position in reciprocal space, mainly due to finite size, strain of the sample and the divergence of the incident beam: $I(\vec{Q}) = I_{hkl} H(\vec{Q} - \vec{Q}_{hkl})$. In this expression, $I(\vec{Q})$ denotes the intensity at a general point $\vec{Q}$ of the reciprocal space, and $H$ is the corresponding spreading function (normalized $\int_{\mathbb{R}^3} H d^3 Q = 1$) and $I_{hkl}$ the integrated intensity (corrected from geometrical factors). Thus, $I_{hkl}$ could be obtained by the following integral in the 3d reciprocal space:

$$I_{hkl} = \iiint_{V_Q} I(\vec{Q}) d^3 Q$$

where $V_Q$ is a volume region around $\vec{Q}_{hkl}$ that must contain the whole diffraction peak.

In contrast, according to the data collection procedure described above, the measured integrated intensity is:

$$I_{hkl}^M = \frac{1}{T}\int_T dt \iint_A dA\, I(\vec{Q})$$

where the first integral refers to the summation during the measurement (scan) and the double one to the summation over the area detector.

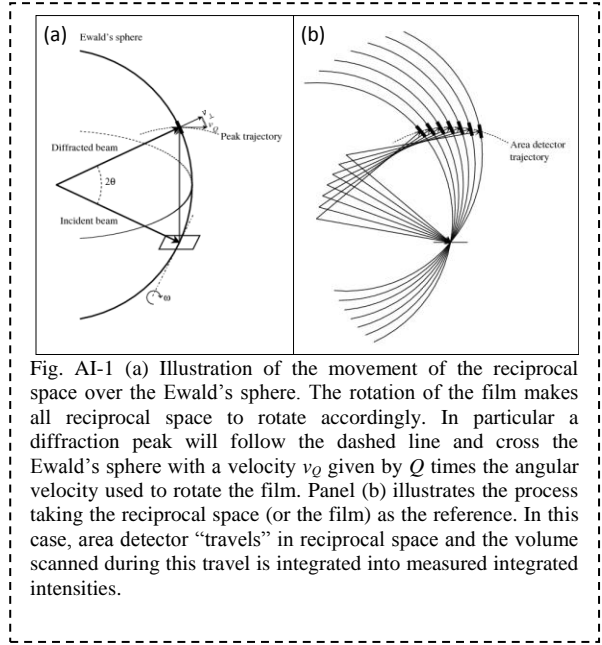

Fig. AI-1 (a) Illustration of the movement of the reciprocal space over the Ewald's sphere. The rotation of the film makes all reciprocal space to rotate accordingly. In particular a diffraction peak will follow the dashed line and cross the Ewald's sphere with a velocity $v_Q$ given by $Q$ times the angular velocity used to rotate the film. Panel (b) illustrates the process taking the reciprocal space (or the film) as the reference. In this case, area detector "travels" in reciprocal space and the volume scanned during this travel is integrated into measured integrated intensities.

Thus, $I_{hkl}^M$ and $I_{hkl}$ are related by the path followed by the diffraction peak to cross the area detector (or, equivalently, the Ewald's sphere) in the reciprocal space due to the film rotation (ω-scan or φ-scan) performed during data collection. More precisely, they are related by the volume element scanned by the detector for the time element, that can be expressed as $d^3Q = dQ_\perp dA = v_\perp dt dA$, where the subscript ⊥ denotes the direction perpendicular to the area detector $A$ and $v\perp$ the projection of the area detector velocity, in reciprocal space, along this direction. This is illustrated, for a ω-scan in Fig AI-1. Velocity of every point in the reciprocal space is given by the angular velocity used for the scan times the distance to the rotation axis. It must be mentioned that this distance, and the velocity, is not homogeneous for all the points covered by the detector, but the variations within it from point to point can be neglected. For a ω-scan this velocity is given by $v_Q = \Omega Q \approx \Omega Q_{hkl} = \Omega \frac{4\pi \sin\theta}{\lambda}$ (Ω is the angular velocity used for the scan) and its projection over the perpendicular to the area detector is given by $v_\perp \approx \Omega \frac{4\pi \sin\theta}{\lambda} \cos\theta = \Omega \frac{2\pi}{\lambda} \sin 2\theta$. This drives to:

$$I_{hkl} = \Delta\omega I_{hkl}^M \frac{2\pi}{\lambda} \sin 2\theta$$

where Δω is the total angle rotated along the ω-scan ($\Delta\omega=\Omega\Delta T$).

For a non-specular reflection, as illustrated in Fig. AI-2 the distance to the rotation axis in reciprocal space is given by $Q \sin \Psi = \frac{4\pi}{\lambda} \sin \theta \sin \Psi$. Thus, the projection of the velocity along the direction perpendicular to the detector (along diffracted beam) is thus given by $v_\perp \approx \Omega \frac{4\pi}{\lambda} \sin \theta \sin \Psi \cos \theta = \Omega \frac{2\pi}{\lambda} \sin 2\theta \sin \Psi$. Akin to the ω-scan, the velocity is not the same for all the point in the detector, but differences are small enough, and this approximation is very reasonable. This expression drives to:

$$I_{hkl} = \Delta\varphi I_{hkl}^M \frac{2\pi}{\lambda} \sin 2\theta \sin \Psi$$

where Δφ is the angle rotated for the φ-scan.

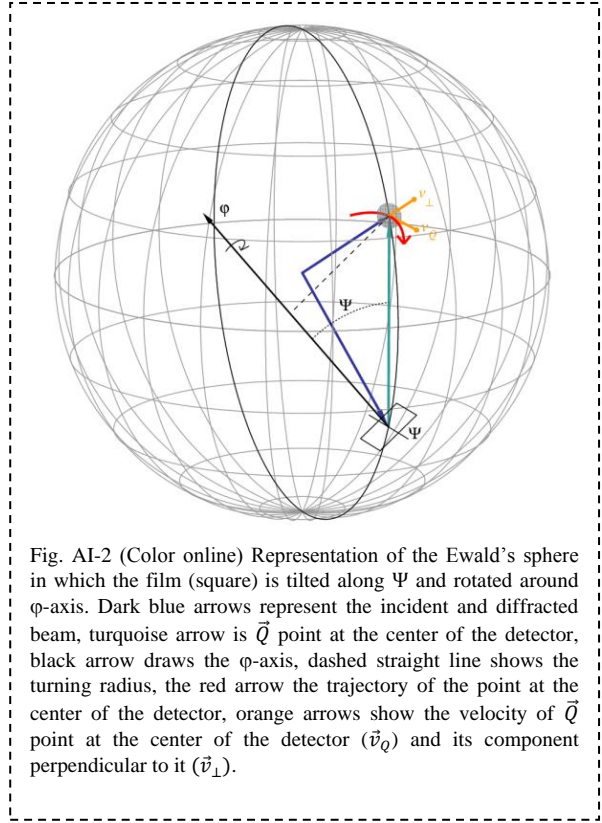

Fig. AI-2 (Color online) Representation of the Ewald's sphere in which the film (square) is tilted along Ψ and rotated around φ-axis. Dark blue arrows represent the incident and diffracted beam, turquoise arrow is $\vec{Q}$ point at the center of the detector, black arrow draws the φ-axis, dashed straight line shows the turning radius, the red arrow the trajectory of the point at the center of the detector, orange arrows show the velocity of $\vec{Q}$ point at the center of the detector ($\vec{v}_Q$) and its component perpendicular to it ($\vec{v}_\perp$).

### III.2 Polarization correction

Integrated intensities are affected by polarization of the beam. The correction is given by the well-known expression [10]:

$$I_{hkl} \propto F_{hkl}^2 \frac{1+\cos^2 2\theta}{2}$$

**III.3 Irradiated volume correction**

Since the sample is a thin film the irradiated volume varies with the angle between the sample and the incident beam ($\gamma$). Diffracted intensity is proportional to the irradiated volume, and this effect must be corrected with:

$$I_{hkl} \propto F_{hkl}^2 V_{irr} = F_{hkl}^2 S \cdot t \frac{1}{\sin \gamma}$$

where $S$ and $t$ are the incident beam section and the film thickness respectively. This expression assumes that the incident beam does not illuminate the whole film. We have checked that this is fulfilled in the case presented in this work using co-planar geometry, but in the case of grazing incidence geometries, this could not be true, and this expression would have to be revised accordingly.

It must also be pointed out that, for specular reflections, the incident angle $\gamma$ is equal to $\omega$ and thus it varies along the scan. Consequently, for this type of reflections, this correction would have to be applied during collection. Nonetheless, we can assume that for the small region (in the reciprocal space) where the diffraction peak is located the variations will not be relevant and a constant $\gamma$ can be employed.

**III.4 Absorption correction**

This correction is due to the fact that direct and diffracted beam are (partially) absorbed by the sample along their path. This absorption is different for every reflection. A beam diffracted at a depth $x$ travels through the sample along a distance given by the expression: $d = x \left[ \frac{1}{\sin \gamma} + \frac{1}{\sin(2\theta - \gamma)} \right]$. Taking this into account and integrating through the whole film the final expression for the absorption correction is:

$$I_{hkl} \propto F_{hkl}^2 \frac{1}{\mu t \left( \frac{1}{\sin \gamma} + \frac{1}{\sin(2\theta - \gamma)} \right)} \left\{ 1 - \exp \left[ -\mu t \left( \frac{1}{\sin \gamma} + \frac{1}{\sin(2\theta - \gamma)} \right) \right] \right\}$$

where $\mu$ is the absorption coefficient of the film and t is the film thickness. If the conditions $\mu t \ll 1$ is satisfied, this correction can be neglected.

**IV Application of $\delta$-recycling and unit cell refinement**

The structure was solved by $\delta$-recycling direct methods [11] as implemented in XLENS_v1 [12]. The fifth set of starting random phases yielded the true solution. The found relative scattering powers and ($x/a$, $y/b$, $z/c$) coordinates were (1000, ¼, ¾, ½) for site 1 (2b Wyckoff position of $P4/nmm$, origin choice 2), (900, ¾, ¾, 0.231) for site 2 and (478, ¾, ¾, 0.823) for site 3 (2c Wyckoff positions). The respective scattering powers strongly suggest that site 1 is fully occupied by Cu, that As is at site 2 and that Mn partially occupies site 3. These results are consistent with independent structural and compositional information presented in the section V.

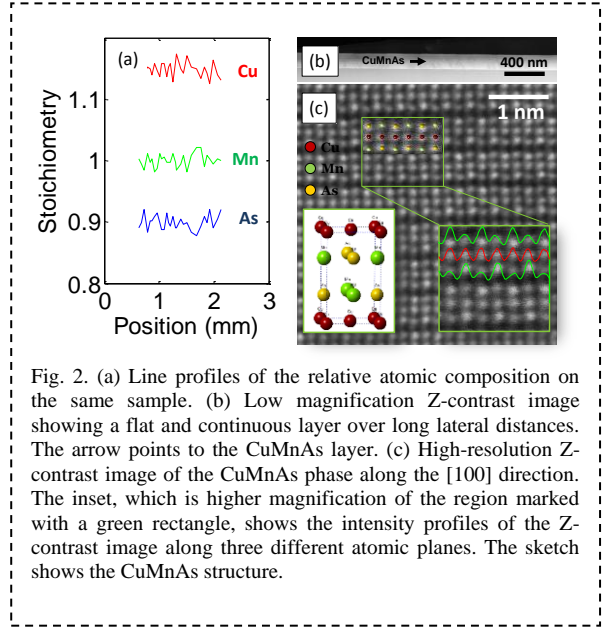

Fig. 2. (a) Line profiles of the relative atomic composition on the same sample. (b) Low magnification Z-contrast image showing a flat and continuous layer over long lateral distances. The arrow points to the CuMnAs layer. (c) High-resolution Z-contrast image of the CuMnAs phase along the [100] direction. The inset, which is higher magnification of the region marked with a green rectangle, shows the intensity profiles of the Z-contrast image along three different atomic planes. The sketch shows the CuMnAs structure.

The $z$ coordinates of sites 2 and 3 together with their respective occupancies were refined with the single-crystal least-squares program SHELX-97 [13]. We employed the following definitions for $wR^2$, $R_1$ and $S$: $wR^2 = \{\Sigma[w(F_o^2-F_c^2)^2]/\Sigma[w(F_o^2)^2]\}^{1/2}$; $R_1 = \Sigma||F_o|-|F_c||/\Sigma|F_o|$; $S= \Sigma\{[w(F_o^2-F_c^2)^2]/(n-p)\}^{1/2}$ with $w=1/\sigma^2(F_o)$, where $n$ is the number of reflections and $p$ the number of refined parameters. The refinement converged to a residual $wR^2 = 0.1770$ for 34 data and 8 refined parameters; $S= 0.643$; $R_1=0.1112$. The values of the $z$ coordinates remained unchanged compared to the direct methods ones [0.236(2), 0.841(5), for sites 2 and 3, respectively] and the individual $U_{eq}$ values were all close to 0.02 Å$^2$. The refined occupancies for sites 2(As) and 3(Mn) are 0.964(15) and 0.863(13), respectively, so that the corresponding sum of electrons in the cell (the three sites) is $1·29 + 0.964·33 + 0.863·25 = 82.39$. This sum, properly scaled, must be equal to the one derived from the atomic proportions measured experimentally. The actual composition of the thin films were obtained by variable voltage electron probe microanalysis, using a CAMECA SX-50 electron microprobe equipped with four wavelength-dispersive x-ray spectrometers. X-ray intensities were measured at 10, 12, 15 and 20 keV electron incident energy and they were analysed with the help of the STRATAGEM package (SAMX, France), which calculates the thickness and composition of a multilayer target by least squares fitting of an analytical x-ray emission model to the experimental data. The results shown in Fig. 2a indicate Cu: 1.13(2), As: 0.88(2),

Mn:0.98(2) and are constant over a distance of several mm well above the beam width of 50 μm employed during the X-ray experiments. Hence, by making $k \cdot (1.13 \cdot 29 + 0.88 \cdot 33 + 0.98 \cdot 25) = 82.39$, it follows that $k=0.9546$. After multiplying by $k$, the atomic content of the cell is 1.08 Cu, 0.84 As and 0.94 Mn. Since site 1 was assumed to be fully occupied by Cu and since from the refinement it is known that there are 0.86 Mn at site 3, this gives for site 2 the composition 0.84 As + 0.08 Cu + 0.08 Mn. The sum is strictly 1.00 and we conclude that the site 2 is essentially fully occupied. It must be highlighted that 0.964 As (the refined occupation) is nearly equivalent (in scattering power) to 0.84 As + 0.08 Cu + 0.08 Mn (31.82 e$^-$ vs. 32.04 e$^-$). In the last refinement cycles due to the limited set of observations we did not attempt the introduction of anisotropic atomic displacement parameters. In conclusion, the refinement converged to (Table II) $wR^2 = 0.1066$ and $R_1=0.0933$ (9 parameters); $S= 0.391$. Respective final $z$ coordinates for sites 2 and 3 are 0.2347(13) and 0.8298(30). The $U_{11}=U_{22}$ and $U_{33}$ atomic displacement parameters are 0.025(6) and 0.084(11) for site 1; 0.013(4) and 0.020(5) for site 2; 0.016(6) and 0.038(11) for site 3.

Table II: Structural details of CuMnAs thin film. Unit cell parameters are $a=3.820(5)$Å and $c=6.318(5)$Å and the space group is $P\,4/nmm$ (no. 129). Cu mainly occupies Wyckoff site 2$b$ (¼, ¾, ½), and Mn and As mainly occupy sites of type 2$c$ (¾ ¾ $z$). The Mn site is not fully occupied and the As one is partially replaced by Cu and Mn. All non-diagonal atomic displacement parameters are zero forced by site symmetry

| 100%-Cu | | 84%-As; 8%-Cu; 8%-Mn | | | 86%-Mn | | |
|---|---|---|---|---|---|---|---|
| $U_{11}=U_{22}$ | $U_{33}$ | $z$ | $U_{11}=U_{22}$ | $U_{33}$ | $z$ | $U_{11}=U_{22}$ | $U_{33}$ |
| 0.025(6) | 0.084(11) | 0.2347(13) | 0.013(4) | 0.020(5) | 0.8298(30) | 0.016(6) | 0.038(11) |

### V Independent structural analyses

For atomic scale information on the structure, we used aberration-corrected STEM. The thin films were examined in a Nion UltraSTEM column, operated at 100 kV and equipped with a fifth order Nion aberration corrector. Specimens for STEM observations were prepared by conventional thinning, grinding, dimpling and Ar ion milling. Figure 2(a) and (b) show a low and a high magnification Z-contrast image of the CuMnAs thin film, respectively. High-resolution images of the interface region, not shown, proved that the interface is sharp, and that in-plane and out-of-plane textured growth occurs. In a STEM, the high-angle annular dark field detector allows recording incoherent Z-contrast images, in which the contrast of an atomic column is approximately proportional to the square of the average atomic number (Z). In this situation, heavier atomic columns can be easily distinguished from lighter ones, as shown in the figure inset, where the intensity profiles of the Z-contrast image along three different atomic planes emphasizes the contrast variation observed in every atomic column. Accordingly, atomic identities can be assigned based on the model obtained from X-ray analyses and the image intensity. In agreement with the conclusions of the X-ray diffraction analyses, the atomic positions replicate the wiggling observed in sites 2 and 3 in tetragonal CuMnAs.

## VI Conclusions

Using only Cu X-ray radiation sources, we have solved and refined the structure of a material prepared in the form of a thin film. Our preliminary X-ray characterizations revealed that the space group of the epitaxy-distorted thin film was different from the bulk compound, despite having the same stoichiometry. With no a priori knowledge we determined the space group by detecting the systematic extinctions in the exhaustive X-ray analyses. We explicitly derived the set of formulas to correct the raw data as a function of the angular coordinates and two scan modes ($\varphi$ and $\omega$-scans) in the standard z-axis geometry and transform the data into a suitable input for single crystal direct methods. The determination and refinement of the unit cell positions concluded in a set of atomic positions and occupancies in excellent agreement with independent STEM and stoichiometry characterizations.

Recently we have applied this methodology to the determination of the bond lengths in a set of $Sr_2IrO_4$ samples with increasing strain finding an excellent agreement with complementary X-ray absorption studies [14]. Our on-going work comprises the study of perfectly epitaxial monolayers and bilayers grown on top of a substrate with very similar lattice parameters. We extend the concept introduced in this work into a dynamical diffraction fitting of all reflections in the unit cell of the constituent layers of the structure. The ability to perform these analyses using common tube sources opens the door to routine in-house studies of strain-induced structure changes, other than lattice distortion, in thin films (for example, atomic positions, bonding angles and distances) and the rich correlation with electrical and magnetic properties this entails. Theory groups could rapidly obtain valuable structural information to guide changes in future growth and characterization experiments.

## Acknowledgements

X.M. acknowledges to Czech Science Foundation (Project P204/11/P339). All authors thank X. Llobet from Scientific and Technological Centers of the University of Barcelona for the chemical analysis. J.G. thanks JAE CSIC grant. Research at UCM (M.A.R.) was supported by the ERC starting investigator award, grant ♯239739 STEMOX. Research supported in part by ORNL's Shared Research Equipment (ShaRE) User Facility, which is sponsored by the Office of BES, U.S. DOE. J.R. and C.F. acknowledge financial support from Spanish Ministerio de Ciencia e Innovación Tecnológica (Projects MAT2009-07967, Consolider NANOSELECT CSD2007-00041) and the Generalitat de Catalunya. T.J. and V.N. acknowledge the support from the ERC Advanced Grant 268066-0MSPIN and from the Ministry of Education of the Czech Republic Grants No. LM2011026. C.R. acknowledges financial support from Fondazione Cariplo via the project EcoMag (Project No. 2010-0584). The work was partially supported by the Czech Science Foundation (project P204/12/0595).